\documentclass[journal=jacsat,manuscript=article]{achemso}

\usepackage{chemformula} 
\usepackage[T1]{fontenc} 

\usepackage{csquotes}
\MakeOuterQuote{"}

\usepackage{siunitx}

\usepackage{bm}
\usepackage{braket}
\renewcommand{\v}[1]{\bm{\mathrm{#1}}}
\newcommand{\m}[1]{\bm{\mathsf{#1}}}

\author{Sangeeta Sharma}
\affiliation[BigPharma]
{Max-Born-Institute for Non-linear Optics and Short Pulse Spectroscopy, Max-Born Strasse 2A, 12489 Berlin, Germany \\
Institute for Theoretical Solid-state Physics, Freie Universit\"at Berlin, Arnimallee 14, 14195 Berlin, Germany}
\author{Deepika Gill}
\affiliation[BigPharma]
{Max-Born-Institute for Non-Linear optics, Max-Born Strasse 2A, 12489 Berlin, Germany}
\author{John Kay Dewhurst}
\affiliation[BigPharma]
{Max-Planck-Institut fur Mikrostrukturphysik Weinberg 2, D-06120 Halle, Germany}
\author{Peter Elliott}
\affiliation[BigPharma]
{Scientific Computing Department, Science and Technology Facilities Council UK Research and Innovation (STFC-UKRI), Rutherford Appleton Laboratory, Harwell Campus, Didcot OX11 0QX, United Kingdom}
\author{Sam Shallcross}
\email{shallcross@mbi-berlin.de}
\affiliation[BigPharma]
{Max-Born-Institute for Non-Linear optics, Max-Born Strasse 2A, 12489 Berlin, Germany}

\title[An \textsf{achemso} demo]
{Ultrafast Saddletronics}

\abbreviations{IR,NMR,UV}

\begin{document}

\begin{tocentry}

\includegraphics[width=0.85\textwidth]{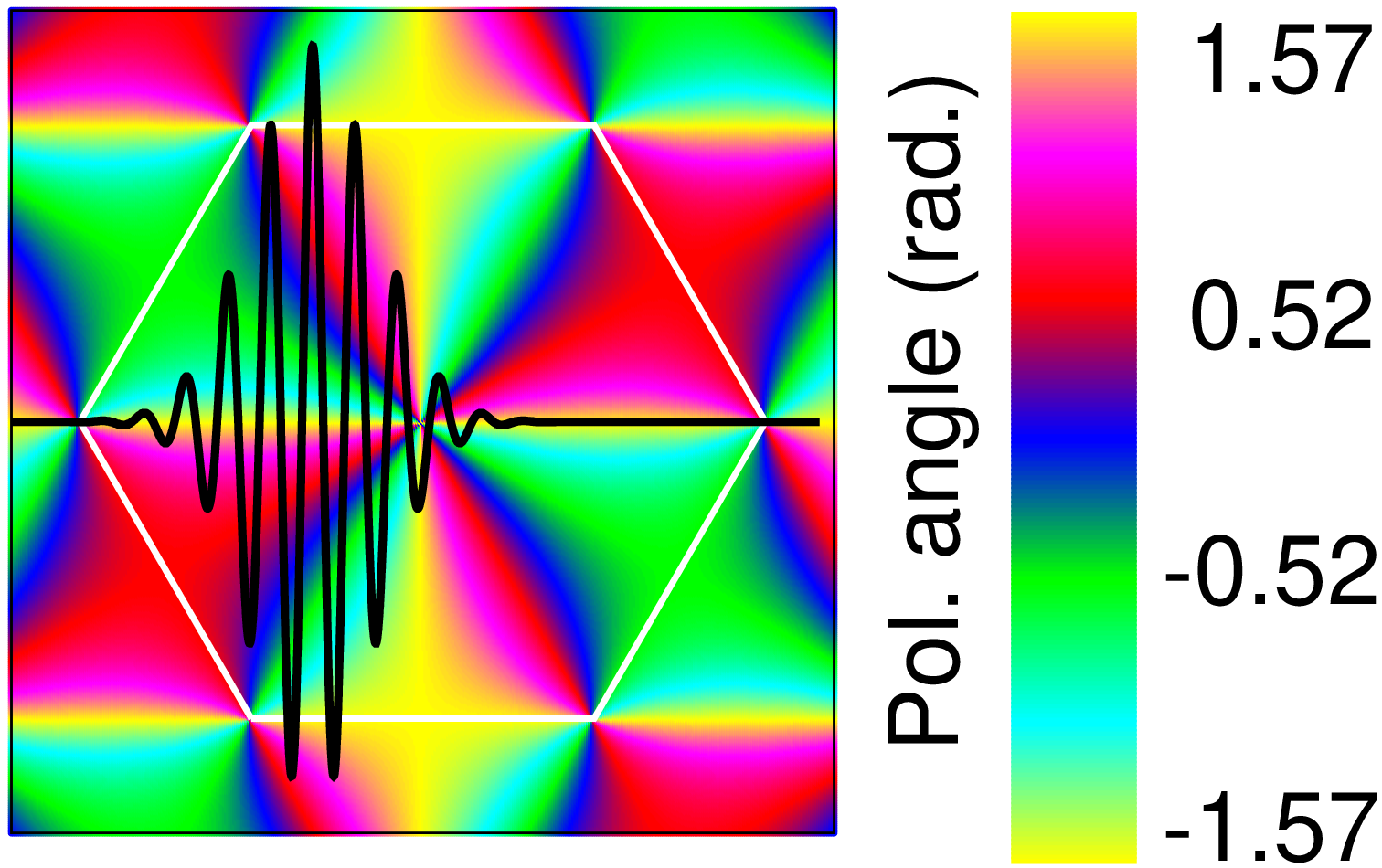}

\end{tocentry}

\begin{abstract}
Low energy valleys in the band structure of 2d materials represent a potential route to the ultrafast writing of information in quantum matter by laser light, with excited charge at the K or K$^\ast$ valleys representing the fundamental states of 1 and 0. Here we demonstrate that a second electronic feature, the saddle point, is endowed with lightwave control over information states. Linearly polarized light is shown to excite 2 of the 3 inequivalent M point saddles in graphene, generating three possible excited configurations, with which of these are realised determined by the polarization vector direction. We show that saddle excitation is highly robust, with "saddle polarized" states created both in the sub-cycle strong field regime and the long time limit of extended multi-cycle pulses. Our findings, applicable to other members of the graphene family and Xenes such as stanene, point towards a rich and ultrafast light based manipulation of matter based on the saddle point.
\end{abstract}

\section{Introduction}

Local energy minima ("valleys") in the electronic conduction bands of certain semiconductors represent an emergent solid state freedom, with control over this freedom the central aim of the field of "valleytronics"\cite{xiao_nonlinear_2015,mak_lightvalley_2018,
langer_lightwave_2018,vitale_valleytronics_2018,
yang_chiral_2019,
li_room-temperature_2020}. Charge excitation at two inequivalent valleys provides a basis for realizing in quantum matter the fundamental information states of "1" and "0"\cite{zibouche_transition-metal_2014,schaibley_valleytronics_2016,
 vitale_valleytronics_2018,zhang_electrically_2021}, with potential application in classical and quantum computing. Laser excitation and control over such states would then manifest as an ultrafast information processing and storage by light. Surprisingly, despite more than a decade of research, there exist very few classes of materials for which such light-matter control exists: certain members of the transition metal dichalcogenide family\cite{xiao_coupled_2012,mak_control_2012,
zibouche_transition-metal_2014,yu_thermally_2015,
xiao_nonlinear_2015,
langer_lightwave_2018,berghauser_inverted_2018,
ishii_optical_2019,yang_chiral_2019,
liu_valleytronics_2019,
silva_all-optical_2022,asgari_unidirectional_2022} and gapped bilayer graphene
\cite{friedlan_valley_2021,yin_tunable_2022}.
Moreover, the band structure feature -- low energy valleys at conjugate K and K$^\ast$ high symmetry points -- remains the same in all cases.

Here we find an unexpected second electronic feature that can be selectively excited by light. The M point saddle in graphene (associated with the low energy van Hove singularity) exhibits ultrafast coupling to linearly polarized light, with pronounced excitation at two of the three inequivalent M points. This generates three information configurations -- as opposed to two in the case of valleytronics -- with the selection of the pair of activated M points determined by the light polarization direction. 

The low symmetry of the M saddle implies that these "saddle states" are innately current carrying, allowing lightwave control over both saddle excitation and saddle current. The physical mechanism underpinning M excitation, quite different from that driving K valley excitation, we show to support the excitation of highly saddle polarized states both in the ultrafast sub-cycle strong field regime and the long time limit of extended multi-cycle pulses. Our results open the way for an ultrafast "saddletronics" of light-matter information manipulation, potentially at times faster than those of quantum decoherence in solids.

\section{Light-saddle coupling}

\begin{figure}[t!]
\includegraphics[width=1.0\textwidth]{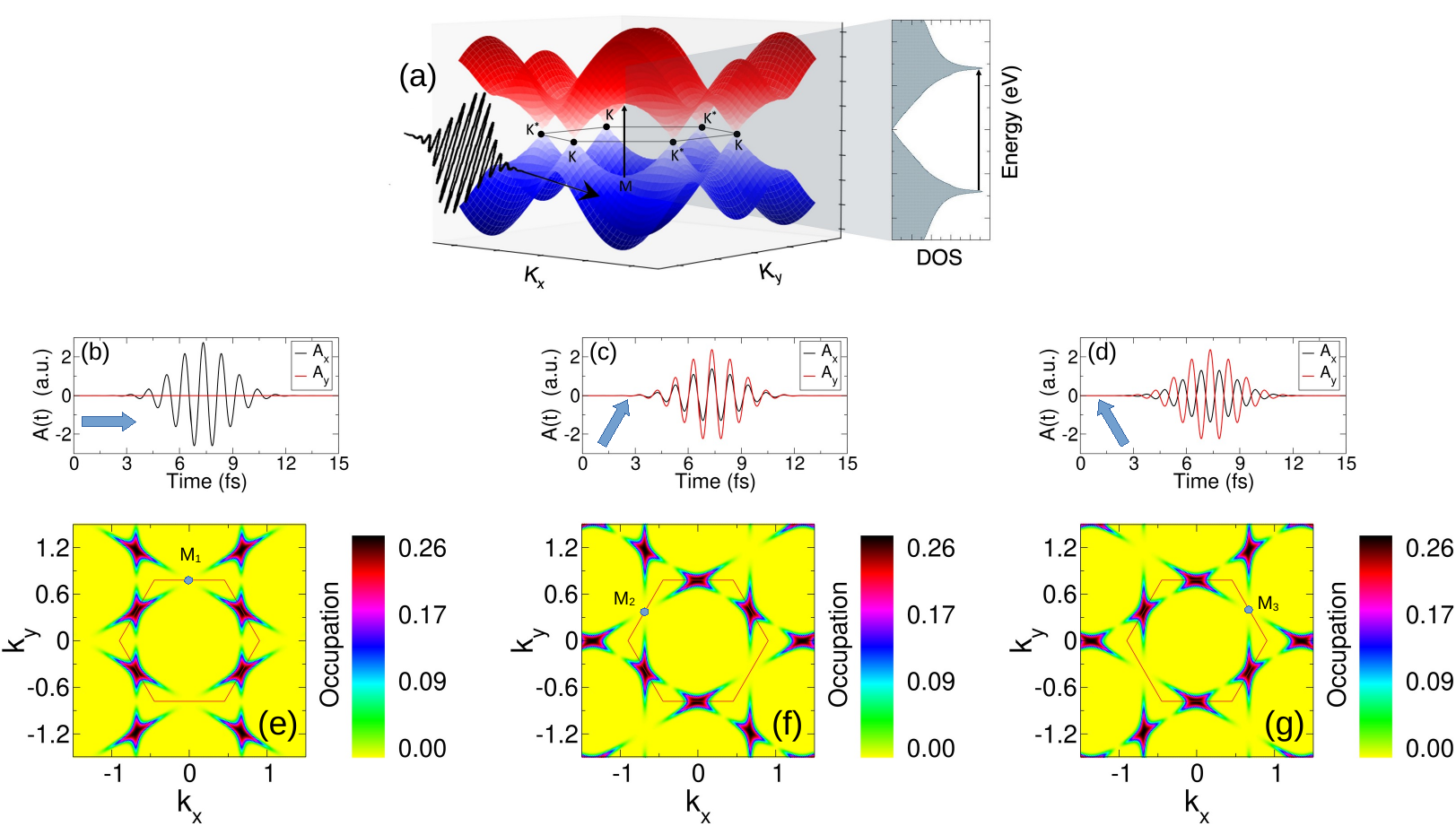}
\caption{\small {\it Light-saddle coupling in graphene}. For a deep ultraviolet light pulse tuned to the M point gap, there exists a profoundly anisotropic response to linearly polarized light. {\bf(a)} Schematic illustration of excitation at the M point, with the saddle structure near M clearly visible. {\bf(b-d)} The vector potential of three linearly polarized light pulses with the polarization vector indicated by the inset arrow, generates, {\bf(e-g)}, a momentum resolved charge excitation at 2 of the three inequivalent M points, with a complete absence of charge at the third, labelled $M_1$, $M_2$, and $M_3$ in (e-g). The polarization direction of saddle tuned light thus determines which configuration of 2 out of 3 inequivalent M points charge is excited at.
}
\label{fig1}
\end{figure}

The graphene Brillouin zone, in addition to the two high symmetry K points, K and K$^\ast$, features three inequivalent high symmetry M points; we label these $M_i$ with $i=1,3$ as shown in Fig.~\ref{fig1}. At each of these the local band manifold features a saddle point in valence and conduction, separated by a 4~eV gap. This implies  activation by a laser pulse frequency in the deep ultraviolet, and a natural question is whether a selection rule holds allowing controlled excitation of individual M points by light.

To investigate light-matter coupling at the M points we employ a dual methodology consisting of both a $\pi$-band only tight-binding model, an approach motivated by the fact that {\it ab-initio} studies find laser excitation largely confined to the $\pi$-band\cite{li_ab_2021}, augmented by full potential time-dependent density function theory to confirm key findings; details of both these approaches are presented in the Methods section.

In Fig.~\ref{fig1}(a) we present a schematic illustration of the saddle point excitation, with in Fig.~\ref{fig1}(b-d) the vector potential of three such linearly polarized light pulses with frequency 4~eV tuned to the M gap. These differ only in the direction of the polarization vectors indicated by the inset arrows. Applied to graphene each of these pulses excites an equal amount of charge but, as can be seen in Fig.~\ref{fig1}(e-g), dramatically different momentum resolved excitations. A localized excitation at two of the M points can be observed, with identically zero  at the third, establishing lightwave control over M point saddle excitation.

In the light-valley coupling that underpins valleytronics -- charge excitation at a gapped K point by circularly polarized light -- the resulting momentum resolved excitation inherits the $C_3$ symmetry of the conduction/valence edge, a situation that yields zero valley current. The "hour glass" charge excitation revealed in Fig.~\ref{fig1}(e-g) is evidently of low symmetry, indicating symmetry lowering from the local band manifold (of $C_2$ symmetry), and suggesting that saddle activation by light will generated inherently current carrying states.

\begin{figure}[t!]
\begin{center}
\includegraphics[width=0.85\textwidth]{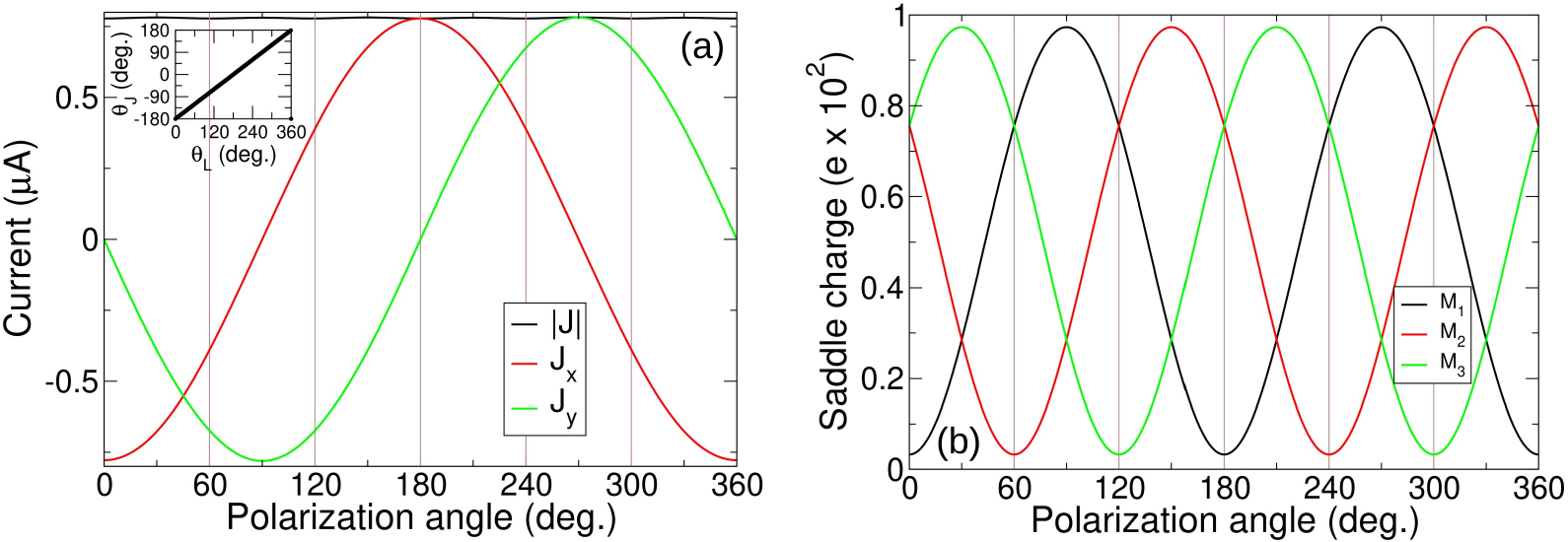}
\caption{\small {\it Light induced "saddle current" in single layer graphene}. Selective excitation of the three inequivalent saddle points is accompanied by a light-controlled current flow. {\bf(a)} The total current as a function of the polarization vector of a linearly polarized pulse (other pulse parameters are identical to the waveforms presented in Fig.~\ref{fig1}(a-c)). As is made clear by the inset panel, the direction of the induced current $\theta_J$ is exactly anti-parallel to the angle of the pulse polarization vector $\theta_L$. Underpinning this is a evolving distribution of the weight of the charge excitation between the three saddle points as the polarization vector changes, panel {\bf(b)}. Note that the total excited charge is identical for all pulses, with only its division between the three saddle points evolving with polarization direction.
}
\label{fig2}
\end{center}
\end{figure}

To probe the "saddle current" induced by light we present in Fig.~\ref{fig2}(a) the total current as a function of the polarization angle of the linearly polarized pulse. The evident sinusoidal variation implies current aligned along the polarization vector of the light pulse, a fact confirmed in a plot of the current angle $\theta_J$ versus the polarization angle $\theta_L$, see inset panel. The resulting THz emission induced by this current thus represents an experimentally measurable signature of underlying momentum space light-saddle coupling; the direction of emitted THz light will be directed exactly opposite to that of the ultraviolet pulse. Rotation of the polarization vector of light redistributes the weight of the charge excitation continuously between the three M points, Fig.~\ref{fig2}(b), each M point being exactly $\pm 60^\circ$ out of phase with the other two M points. The three "principle" saddle configurations shown in Fig.~\ref{fig1} correspond to the cases $0^\circ$, $60^\circ$, and $120^\circ$, at which the charge at $M_1$, $M_2$, and $M_3$ respectively falls to near zero. 

This behaviour strongly suggests the existence of an underlying M point selection rule involving linearly polarized light, similar to the K point selection rule that governs the coupling of the helicity of circularly polarized light to the valley index. Calculation of the matrix element for excitation across the gap at the M points of graphene (see Methods) reveals a probability of excitation $T(\omega)$ given by

\begin{equation}
T(\omega) = \frac{32 \pi v_F^2}{9} (\v M_i.\v A_0)^2 \delta(\varepsilon_M - \omega)
\label{MT}
\end{equation}
where $\varepsilon_M=4$~eV and $\v M_i$ are the M point gap and the crystal momenta respectively (with the index $i=1,3$ denoting the three inequivalent M points), and $\omega$ and $\v A_0$ the frequency and polarization vector of the incident light; $v_F$ the Fermi velocity of graphene. From Eq.~\ref{MT} we see that for a light pulse polarization vector perpendicular to the crystal momentum $\v M_i$ excitation is forbidden, implying three directions of light at which only two of the three M points will be excited, the behaviour seen in Fig.~\ref{fig1}. This result establishes a direct connection between Brillouin zone geometry, encoded via the $\v M_i$ vectors, and light pulse excitation expressed through the polarization vector $\v A_0$. From Eq.~\ref{MT} we also see that the $\pm 60^\circ$ rotation between different M points implies saddle charge excitations that will be precisely $\pm 60^\circ$ out of phase with each other, exactly as seen in Fig.~\ref{fig2}.

Having established the existence of charge and current excitation at saddle points, we now consider their lifetime. Valley excitation persists on the scale of several picoseconds\cite{pogna_hot-carrier_2021}, and while the higher energy of the saddle may offer faster de-excitation we note that scattering from M to K will involve a large momentum. Large momentum intervalley scattering times in graphene are found over a wide range of 1-6 picoseconds\cite{zollner_graphene_2021,
couto_random_2014,zihlmann_out--plane_2020,vasileva_strongly_2017}, suggesting long lived picosecond saddle excitations. This however, being strongly influenced by the environment of the graphene as well as its intrinsic disorder and, being sample dependent, must ultimately be tested by experiment.

\section{Robustness of light-saddle excitation}

\begin{figure}[t!]
\begin{center}
\includegraphics[width=1.0\textwidth]{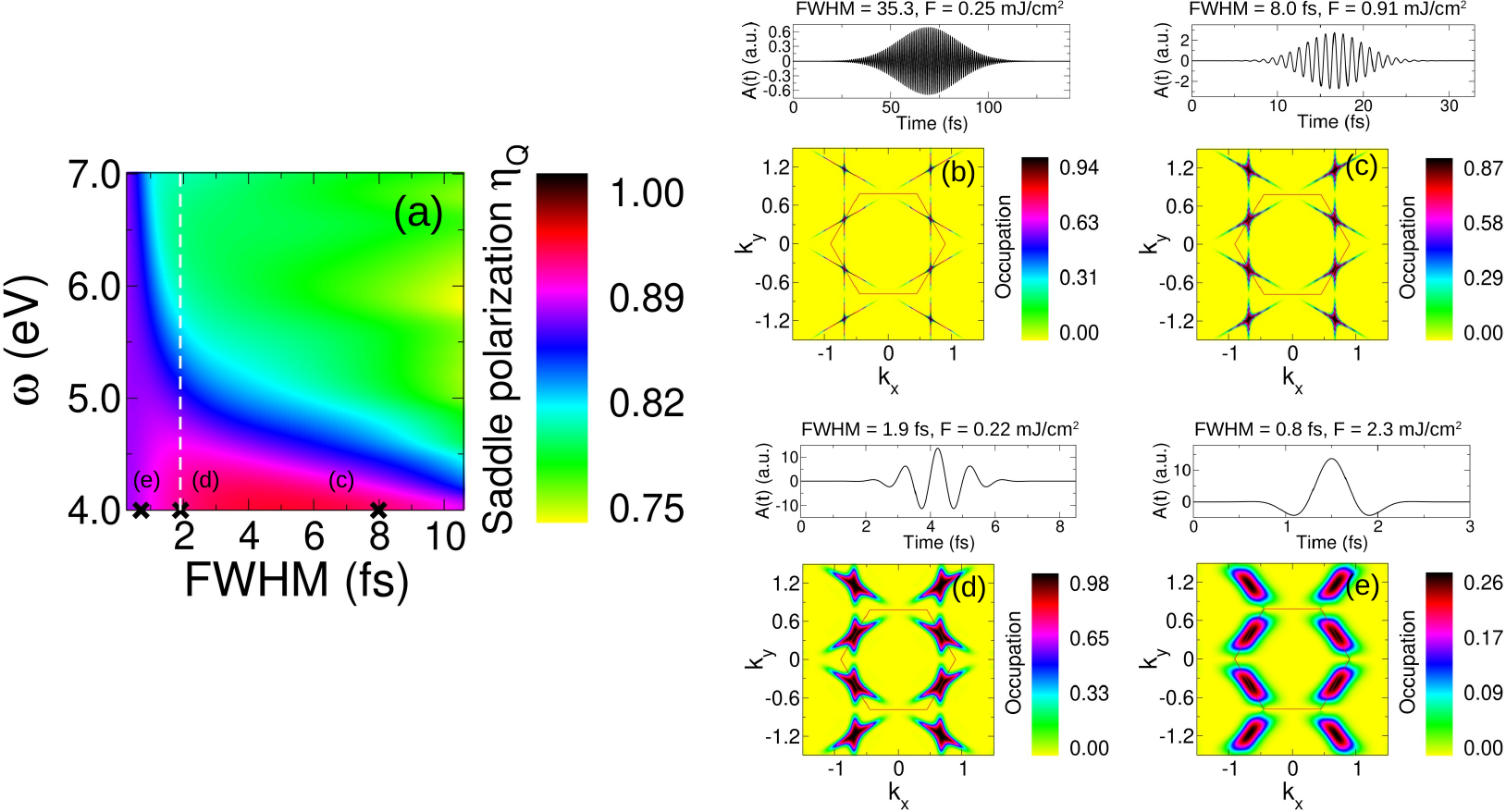}
\caption{\small {\it Light-saddle coupling map: saddle polarization explored over pulse parameters}. {\bf(a)}  Saddle polarization, defined as the normed difference of the saddle charges at $M_1$ and $M_2$, $\eta_M = (Q_{M_1} - Q_{M_2})/(Q_{M_1} + Q_{M_2})$, is presented as a function of pulse duration and energy revealing that nearly perfect saddle polarization exists both for longer time pulses and in the ultrafast single cycle regime. The polarization vector is taken to be perpendicular to the $M_1$ special point, see Fig.~\ref{fig1}(a). The vertical dashed line represents the current "world record" for a deep ultraviolet pulse of energy 4~eV\cite{galli_generation_2019}. {\bf (b-e)} Momentum resolved excitation for three representative cases as indicated in panel (a). At long pulse duration the saddle excitation presents a highly localized excitation at the M points, that in the short pulse limit broadens around the M point, with reduction in amplitude of the excited charge. For both short and long times near complete saddle polarization is seen, with charge excited at $M_2$ and $M_3$ but not at $M_1$. The amplitude of the pulse employed here is 0.02~a.u., however similar findings are found for any sensible variation of light pulse amplitude.
}
\label{fig3}
\end{center}
\end{figure}

We now consider the temporal robustness of light-saddle coupling, addressing both the sub-cycle as well as long time limits. To this end we define, in analogy to the case of valley polarization, a "saddle polarization" parameter. Without loss of generality we take the lightwave polarization vector to have angle $0^\circ$. In this case, see Fig.~\ref{fig1}, we may then define saddle polarization as the normed difference between the saddles charges at $M_1$ and $M_2$: $\eta_M = (Q_{M_1} - Q_{M_2})/(Q_{M_1} + Q_{M_2})$. A saddle polarization of $\eta_M=+1$ then indicates the fully polarized situation shown in Fig.~\ref{fig1}e, with $\eta_M=0$ corresponding to complete loss of saddle polarization.

At the ultrafast limit high saddle polarization persists over a wide range of pulse durations and frequency, Fig.~\ref{fig3}(a), with pulse durations well into the sub-cycle attosecond regime seen to preserve saddle polarization. On the other hand at long duration, Fig.~\ref{fig3}(b), and well defined pulse central frequency a highly saddle polarized state ($\eta_M = 0.86$) is still achieved, strongly localized at the M point. Thus both the sub-single cycle and long time limits light-saddle coupling is effective. As the saddle point reduces from the long time to the single cycle limits this highly localized excitation broadens, Fig.~\ref{fig3}(c-e), with the pulses and momentum resolved excitations corresponding to those points labelled (c-e) in panel (a). The 1.9~fs pulse, panel (d), represents the current short time "world record" for a deep ultraviolet pulse.

Underlying this temporal persistence is that the M point transmission, Eq.~\ref{MT}, holds not only at the M points but also also along the M-K and M-$\Gamma$ directions, see Methods and also Supplemental in which the light-matter coupling is derived. Broadband short time pulses that excite charge across a range of energies close to the saddle point thus follow the same light-matter coupling as long time pulses whose well defined central frequency excites charge only at the M point. Similarly large amplitude pulses -- whose dynamical evolution involves crystal momenta far from the M points -- behave exactly as pulses whose dynamical evolutions remains close to M. The pulse behaviour shown Fig.~\ref{fig3} is thus found at a wide range of amplitudes, see Supplemental. Light-saddle coupling is thus seen to be extremely robust, and both ultrashort large amplitude pulses, and weak amplitude long time pulses, allow full control over saddle polarization.

\section{{\it Ab-initio} treatment of light-saddle coupling}

\begin{figure}[t!]
\includegraphics[width=0.95\textwidth]{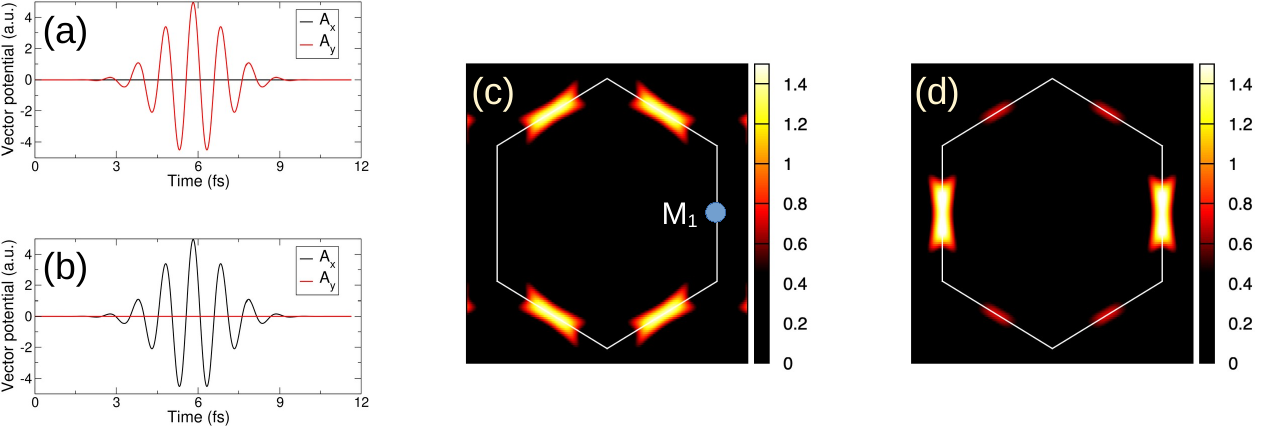}
\caption{\small {\it Density functional treatment of light-saddle coupling}. Two principle cases of light-saddle coupling treated via time-dependent density functional theory: the polarization vector parallel and perpendicular to the crystal momenta of one of the M points. {\bf (a,b)} M point gap tuned linearly polarized pulses with perpendicular and parallel polarization vector to the high symmetry point labelled $M_1$ in (c). These two pulses generate {\bf (c)} zero charge excitation at $M_1$ and {\bf (d)} predominant excitation at $M_1$ with $\sim 1/3$ reduced intensity at the two remaining M points.
}
\label{fig4}
\end{figure}

Having established light-saddle coupling on the basis of $\pi$-band tight-binding simulations we now extend our treatment to full potential time-dependent density function theory calculations. This introduces both an accurate description of the ground state band structure, but also a fully time dependent description of the evolving wavefunction. In Fig.~\ref{fig4} we show two key cases that illustrate the physics of M point excitation as encoded in Eq.~\ref{MT}: a gap tuned linearly polarized pulse with polarization vector perpendicular and parallel to one of the M points, panels (a) and (b) respectively. The former case should, on the basis of Eq.~\ref{MT}, yield zero excitation, with the latter maximal excitation at the M point collinear with the pulse polarization vector and an intensity reduced by $1/4$ at the remaining two. 

The momentum resolved excitation corresponding to (a) and (b) exhibit precisely the physics expected on the basis of the tight-binding analysis. With the polarization vector aligned perpendicular to $M_1$, the $x$-directed vector potential shown in panel (a), zero charge is excited at this point, panel (c). In contrast for the $y$-directed vector potential, that is parallel to $M_1$, charge is excited preferential at that M point, with a reduction in intensity of $\sim 1/3$ at the remaining two, panel (d).

It is also notable that for graphene, as the $\pi$-manifold dominates the region near the Fermi energy (the all-electron band structure may be found in the Supplemental), excitation at the saddle point does not induce other "stray" transitions in the Brillouin zone, as confirmed by the saddle localized excitation seen Fig.~\ref{fig4}(c,d). While this situation guaranteed to hold for other members of the graphene family, for example rhombohedral and twist few layer graphenes, it may not be the case in other Xenes that possess significant $\pi$-$sp^2$ hybridization arising from their buckled structure.

\section{Discussion}

Coupling between distinct band manifold features, such as "valleys", and the vector potential structure of an ultrafast laser pulse underpin the drive to control the solid state by light. Here we have shown that M point tuned light pulses in graphene excite charge according to a simple geomantic rule: the vector product of the polarization vector and the M point crystal momentum determines the probability of excitation. The allows three configurations at which 2 out of the 3 inequivalent M points possess localized charge excitation, with identically zero at the third. In any electronics one requires not only encoding of "bits" via selective excitation of regions of the Brillouin zone, but also transform of information via current flow. Light excited saddle states, in contrast to valley excitations, are intrinsically endowed with current. The angle of THz emission we predict to be collinear with angle of saddle tuned light pulse, proving an experimental test of the underlying momentum space light-saddle coupling.

While we have demonstrated light-saddle coupling for single layer graphene, this physics will be valid for the entire graphene family, including few layer and twisted graphenes, both of which have no gap but will allow control by light via the saddle point. Other Xenes, stanene, germanene, and the silicenes, also represent potential materials for an ultrafast control of saddle points via light, with the possibility of including spin physics in the heavy atom Xenes. There thus appears rich possibilities for a 2d material "saddletronics" in which the saddle point plays the role of the manifold feature addressed by ultrafast laser pulses.

\section{Methods}

\subsection{Tight-binding calculations}

{\it Laser pulse}: We employ a Gaussian envelope centred at $t_0$:

\begin{equation}
\v A(t) = \v A_0 \exp\left(-\frac{(t-t_0)^2}{2\sigma^2}\right) \sin\omega (t-t_0)
\label{pulse}
\end{equation}
where $\v A_0$ is the pulse polarization vector (with the pulse amplitude $|\v A_0|$), $\sigma$ is related to the full width half maximum by $FWHM = 2\sqrt{2\ln{2}}\sigma$, and is $\omega$ the central frequency of the light pulse. This form is used for both the tight-binding and time-dependent density functional theory calculation.

{\it Time propagation within the tight-binding approach}: The initial state is provided by a Fermi-Dirac distribution with $T=0$, and we expand the time-dependent wavefunction at crystal momenta $\v k$ using Bloch states $\ket{\Phi_{\alpha \v k}}$ where $\alpha=1,2$ labels the two sub-lattices of graphene:

\begin{equation}
\ket{\Psi_{\v q}(t)} = \sum_{\alpha} c_{\alpha \v q}(t) \ket{\Phi_{\alpha \v k(t)}}
\label{eq:S1}
\end{equation}
The time-dependent Schr\"odinger equation can be written in terms of the column vector of these time dependent expansion co-efficients as

\begin{equation}
i \partial_t c_{\v q}(t) = H(\v k(t)) c_{\v q}(t)
\label{eq:cSE}
\end{equation}
where $\v k(t)$ is given by the Bloch acceleration theorem

\begin{equation}
\v k(t) = \v k(0) - \v A(t)/c
\end{equation}
and $H(\v k(t))$ is the Hamiltonian of graphene in the $\pi$-band only approximation and expressed in the Bloch basis:

\begin{equation}
H(\v k) = \begin{pmatrix}
0 & t_{\v k} \\
t_{\v k}^\ast & 0
\end{pmatrix}
\label{h2}
\end{equation}
with the Bloch sum $t_{\v k} = -t \sum_j e^{i \v k.\v r_j}$ where $t=-2.0$~eV is the nearest neighbour hopping and $\{\v r_j\}$ the nearest neighbour vectors of the honeycomb lattice.

For the numerical propagation of Eq.~\ref{eq:cSE} we employ the Crank-Nicolson method. A $300\times 300$ k-mesh is employed with a time step of 20 attoseconds. 

\subsection{Light-saddle coupling}

To derive a selection rule for the coupling of light to the M point saddle we frame the following question: what is the polarization vector generating maximal excitation at each crystal momenta in the first Brillouin zone of graphene? The
Hamiltonian for light-matter coupling can be written in velocity gauge as

\begin{equation}
H(\v k) = H_0(\v k) + \m\nabla_{\v k} H_0(\v k) . \v A
\label{gH}
\end{equation}
where the velocity operator is given by

\begin{figure}[t!]
\includegraphics[width=0.45\textwidth]{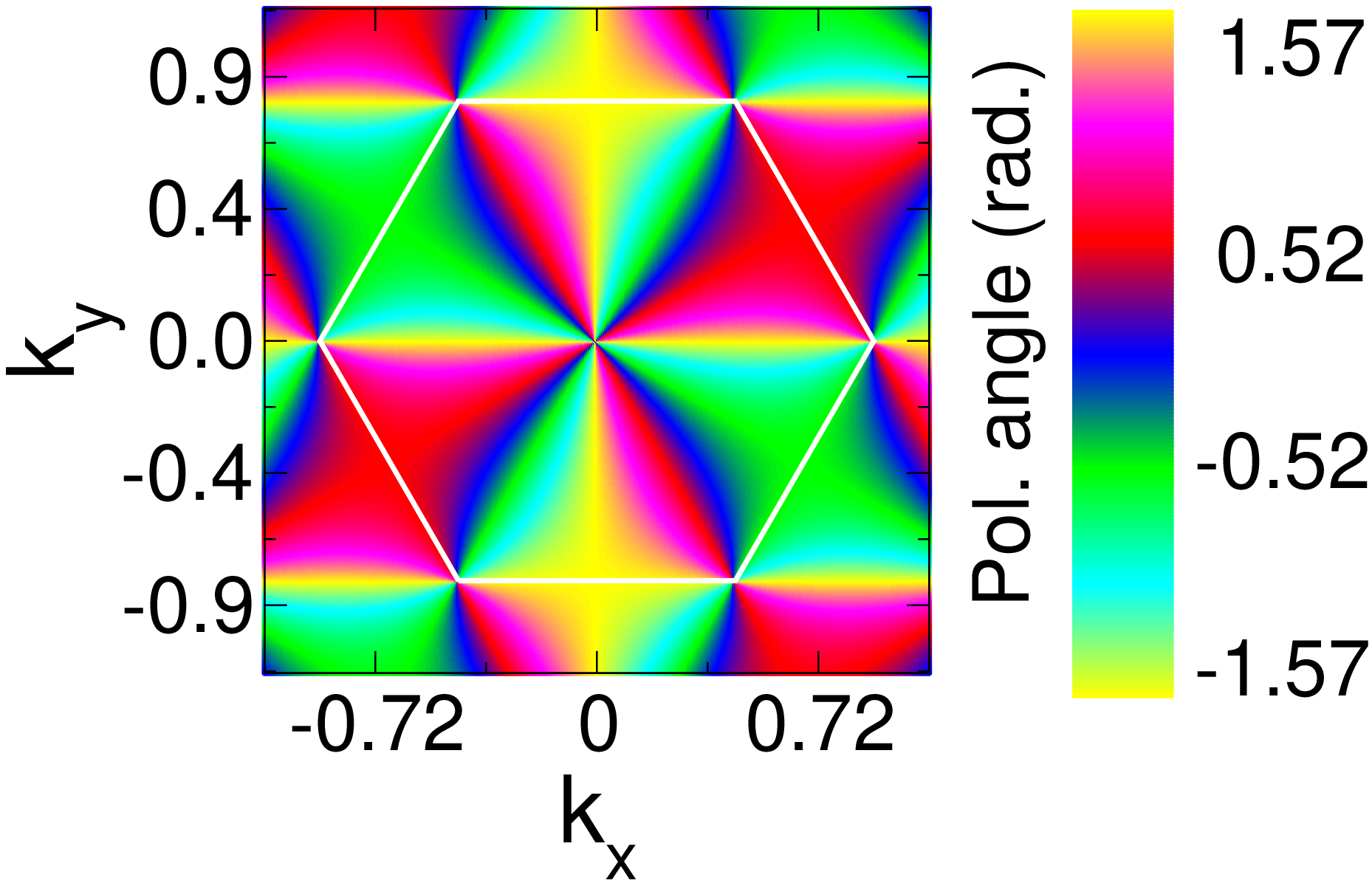}
\caption{\small {\it Polarization vector generating maximal coupling of linearly polarized light in graphene.}. The excited charge at each crystal momentum $\v k$ is given by $Q \propto (\v u(\v k).\v A_0)^2 \delta (2\varepsilon_{\v k} - \omega)$, where $\v A_0$ is the polarization vector the linearly polarised light with energy equal to the gap at $\v k$, $2\varepsilon_{\v k}$ and the angle of vector $\v u(\v k)$ is presented here as as a function of momentum $\v k$. When $\v A_0 \parallel \v u(\v k)$ one evidently has maximal light-matter coupling at $\v k$. Note that $\v u(\v k)$ changes slowly in the vicinity of the M points.
}
\label{fig5}
\end{figure}

\begin{equation}
\m\nabla_{\v k} H_0 = \begin{pmatrix}
0 & \m\nabla_{\v k} t_{\v k} \\
\m\nabla_{\v k} t_{\v k}^\ast & 0
\end{pmatrix}.
\end{equation}

We now consider a harmonic perturbation, which can be expressed as $\v A = \v A^{(+)} e^{i\omega t} + \v A^{(-)} e^{-i\omega t}$, in which $A^{\pm}$ are the excitation (+) and de-excitation (-) components of the light wave.
Excitation from valence (v) to conduction (c) in linear response given by

\begin{equation}
T(\v k) = 2\pi \left| \bra{a_{\v k,c}} \m\nabla_{\v k} H_0 \ket{a_{\v k,v}}.\v A^{(+)}\right|^2
\delta(2\epsilon_{\v k} - \omega)
\label{eqT}
\end{equation}
where $2\epsilon_{\v k}$ is the gap between conduction and valence at crystal momenta $\v k$. Employing the
eigenvalues $\varepsilon_{\v k,c/v} = \pm v_F k$
and eigenvectors of the graphene Hamiltonian Eq.~\ref{gH}

\begin{equation}
\ket{a_{\v k c/v}} = \frac{1}{\sqrt{2}} \begin{pmatrix} 1 \\ \pm  e^{i\theta} \end{pmatrix},
\end{equation}
where in each case the + sign refers to the conduction (c) band and the - sign to the valence (v) band, the transition probability Eq.~\ref{eqT}
can be evaluated as

\begin{equation}
T(\v k) = 2\pi\left|\Im (e^{i\theta}  \m\nabla_{\v k} t_{\v k})
 .\v A^{(+)}
 \right|^2.
\end{equation}
This yields immediately for the vector $A^{+}$ the $\v k$-dependent polarization direction

\begin{eqnarray}
\v u(\v k) & = & \Im e^{i\theta} \m\nabla_{\v k} t_{\v k}
\end{eqnarray}
plotted in Fig.~\ref{fig5} over the first Brillouin zone of graphene.

\subsection{Time-dependent density functional theory calculations}

Real-time time dependent density functional theory (TD-DFT)\cite{runge1984,sharma2014} rigorously maps the computationally intractable problem of interacting electrons to a Kohn-Sham (KS) system of non-interacting electrons in an effective potential. The time-dependent KS equation is:
\begin{align}  
\begin{split}
i \frac{\partial \psi_{j}({\bf r},t)}{\partial t} =
\Bigg[
\frac{1}{2}\Big(-i{\nabla}&-\frac{1}{c}\big({\bf A}(t)+{\bf A}_{\rm xc}(t)\big)\Big)^2 
+ v_{s}({\bf r},t) \Bigg]
\psi_{j}({\bf r},t),
\end{split}
\label{e:TDKS}
\end{align}
where $\psi_j$ is a KS orbital and the effective KS potential $v_{s}({\bf r},t) = v({\bf r},t)+v_{\rm H}({\bf r},t)+v_{\rm xc}({\bf r},t)$ consists of the external potential $v$, the classical electrostatic Hartree potential $v_{\rm H}$ and the exchange-correlation (XC) potential $v_{\rm xc}$. The vector potential ${\bf A}(t)$ represents the applied laser field within the dipole approximation (i.e., the spatial dependence of the vector potential is absent) and ${\bf A}_{\rm xc}(t)$ the XC vector potential.

{\it Computational parameters for the TD-DFT calculations}: We employ a $30\times 30 \times 1$ k-mesh, 75 empty states corresponding to a energy cutoff of 70~eV, and the adiabatic local density approximation (LDA) as our exchange correlation functional $v_{xc}$. A time step is 2.4 attoseconds is employed in the time propagation. The electronic temperature is set to 300~K. The unit cell dimensions are $a=b=2.41$~\si{\angstrom} and $c=20.0$~\si{\angstrom}.

\begin{acknowledgement}

Sharma would like to thank DFG for funding through project-ID 328545488 TRR227 (projects A04), and Shallcross would like to thank DFG for funding through project-ID 522036409 SH 498/7-1. Sharma and Shallcross would like to thank the Leibniz Professorin Program (SAW P118/2021). The authors acknowledge the North-German Supercomputing Alliance (HLRN) for providing HPC resources that have contributed to the research results reported in this paper.

\end{acknowledgement}

\begin{suppinfo}

Supporting information is available:
\begin{itemize}
  \item Filename: SI.pdf
\end{itemize}

\end{suppinfo}


\providecommand{\latin}[1]{#1}
\makeatletter
\providecommand{\doi}
  {\begingroup\let\do\@makeother\dospecials
  \catcode`\{=1 \catcode`\}=2 \doi@aux}
\providecommand{\doi@aux}[1]{\endgroup\texttt{#1}}
\makeatother
\providecommand*\mcitethebibliography{\thebibliography}
\csname @ifundefined\endcsname{endmcitethebibliography}
  {\let\endmcitethebibliography\endthebibliography}{}

\end{document}